\documentclass[conference]{IEEEtran}
\usepackage{graphicx,subfigure}
\usepackage{amsmath,amsthm,amssymb}
\usepackage{algorithm,algorithmic}
\usepackage{xcolor}
\usepackage{cite}
\usepackage{subfigure}

\begin{document}

\title{Performance Evaluation of Channel Decoding With Deep Neural Networks}
\author{\IEEEauthorblockN{Wei~Lyu, Zhaoyang~Zhang\IEEEauthorrefmark{2}, Chunxu~Jiao, Kangjian~Qin, and Huazi~Zhang}
	\IEEEauthorblockA{College of Information Science and Electronic Engineering, Zhejiang University, Hangzhou, China.\\
		Provincial Key Lab of Info. Proc., Commun. \& Netw. (IPCAN), Zhejiang, China.\\
		E-mails: \IEEEauthorrefmark{2}ning\_ming@zju.edu.cn}}
\maketitle

\begin{abstract}
With the demand of high data rate and low latency in fifth generation (5G), deep neural network decoder (NND) has become a promising candidate due to its capability of \emph{one-shot} decoding and parallel computing. In this paper, three types of NND, i.e., multi-layer perceptron (MLP), convolution neural network (CNN) and recurrent neural network (RNN), are proposed with the same parameter magnitude. The performance of these deep neural networks are evaluated through extensive simulation. Numerical results show that RNN has the best decoding performance, yet at the price of the highest computational overhead. Moreover, we find there exists a \emph{saturation length} for each type of neural network, which is caused by their restricted learning abilities.

\end{abstract}

\section{Introduction} \label{introduction}

Recently, the long-term evolution, also known as fifth generation (5G), has been widely rolled out in many countries. There is no doubt that 5G can accommodate the rapid increase of user data and system capacity. Intuitively, higher transmission rate requires lower decoding latency. However, conventional decoding algorithms suffer from high decoding complexity that involves a number of iterative calculations. As such, designing new high-speed low-latency decoders has become an emerging issue to be coped with.

Recent advances in deep learning provide a new direction to tackle this problem. Deep learning \cite{LeCun2015} has been applied in computer vision \cite{He2016}, natural language processing \cite{Sutskever2014}, autonomous vehicles \cite{Chen2015} and many other areas. The remarkable results verify its good performance. Inspired by this, the general decoding problem can be viewed as a form of classification, which is a typical application of deep learning. In brief, the deep neural network uses a cascade of multiple layers with nonlinear processing units to extract and transform features contained in encoding structure and noise characteristic. Compared with conventional iterative decoding, the deep neural network decoder (NND) calculates its estimator by passing each layer only once with the pre-trained neural network, which is referred to as \emph{one-shot} decoding. It provides a foundation for low-latency implementations. In addition, the high-speed demand can be easily satisfied by utilizing the current deep learning platforms, such as Tensorflow \cite{Abadi2016}. Generally, they support parallel computing and exploit the powerful hardwares like graphical processing units (GPUs).

Researchers have tried to solve channel decoding problem using deep neural network. The authors in \cite{Nachmani2016} have demonstrated that by assigning proper weights to the passing messages in the Tanner graph, comparable decoding performance can be achieved with less iterations than traditional belief propagation (BP) decoders. These weights are obtained via training in deep learning, which partially compensates for the effect of small cycles in the Tanner graph. Considering that BP decoding contains many multiplications, \cite{Lugosch2017} proposed a light-weight neural offset min-sum decoding algorithm, with no multiplication and simple hardware implementation. \cite{Gruber2017} found that the structured codes are indeed easier to learn than random codes, and addressed the challenge that deep learning based decoders are difficult to train for long codes. Consequently, \cite{Cammerer2017} proposed to divide the polar coding graph into sub-blocks, with the decoder for each sub-codeword being trained separately.

Although the combination of channel decoding and deep neural network has been studied in the above works, two important problems have not been fully investigated. First, which type of deep neural network is more suitable for NND. Second, how the length of codeword affects the NND performance. In this paper, three types of NND, which build upon multi-layer perceptron (MLP), convolution neural network (CNN) and recurrent neural network (RNN), are proposed with the same parameter magnitude. We compare the performance among these three deep neural networks through numerical simulation, and find that the RNN has the best decoding performance at the price of the highest computational overhead. Also, we find the length of codeword influences the fitting of deep neural network (overfitting and underfitting). It is inferred that there exists a \emph{saturation length} for each type of deep neural network, which is caused by their restricted learning abilities.

The rest of this paper is organized as follows. In Section \ref{system design}, the system framework and the proposed structures of NND are provided. Section \ref{performance evaluation} shows the numerical results and provides the comparisons among MLP, CNN and RNN that are trained with and without noise. Section \ref{conclusion} concludes this paper.

\section{System design} \label{system design}
In order to better present the design of NND, we first describe the system framework of NND. Specifically, the training process of NND is introduced in detail and the reason why we set some parameters like training ratio of codebook set is explained. Finally, we describe the proposed structures of MLP, CNN and RNN, respectively.

\subsection{System Framework}

\begin{figure}[!t]
	\center
	\includegraphics[scale=0.31]{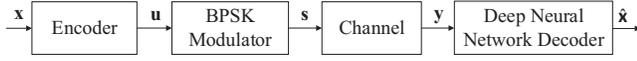}
	\caption{The architecture of deep neural network decoder.}
	\label{framework}
\end{figure}

The architecture of NND is illustrated in Fig.\,\ref{framework}. At the transmitter, we assume that the length of information bits $\textbf{x}$ is $K$. Then, $\textbf{x}$ is encoded to a binary codeword $\textbf{u}$ of length $N$ through a channel encoder, and the codeword $\textbf{u}$ is then mapped to a symbol vector $\textbf{s}$ through the binary phase shift keying (BPSK) modulation. It is assumed that the BPSK symbols are transmitted in the additive white Gaussian noise (AWGN) in this paper.

At the receiver, vector $\textbf{y}$ is received and can be written as

\begin{equation}
\textbf{y} = \textbf{s} + \textbf{n},
\end{equation}
where $\textbf{n} \sim N(\textbf{0}, \sigma^2\textbf{I}_N)$ represents the $N \times 1$ symbol vector.

The estimated information bits $\hat{\textbf{x}}$ is decoded from $\textbf{y}$ with the aid of the NND, where the structure of NND substantially affects the performance. For the notational convenience of the following paper, we denote that $\hat{\textbf{x}}\triangleq\left[\hat{x}_{0},\dots,\hat{x}_{K-1}\right], \textbf{x}\triangleq\left[x_{0},\dots,x_{K-1}\right], \textbf{y}\triangleq\left[y_{0},\dots,y_{N-1}\right]$, and we refer to the set of all possible $\textbf{x}$ and $\hat{\textbf{x}}$ as $\mathcal{X}$  and that of all possible $\textbf{y}$ as $\mathcal{Y}$. Generally, the aim of decoding algorithm is to find a optimal map function $f^* : \mathcal{Y} \rightarrow \mathcal{X}$, which satisfies the maximum a posteriori (MAP) criterion

\begin{equation}
f^*(\textbf{y}) = \mathop{\arg\max}_{\mathbf{x} \in \mathcal{X}} P(\mathbf{x} | \mathbf{y}).
\label{map}
\end{equation}

Obviously, we hope that the NND can reach the performance of MAP decoding as far as possible. As a supervised learning method, the construction of neural network needs two phases, they are training phase and testing phase, respectively. During the training phase, a number of training samples are used to correct the weights and biases in the neural network with the aim of minimizing the loss function, and the map function $f$ will be obtained after this phase. Then, the testing phase, which is the actual decoding phase, is just to estimate information bits from the new received symbol vector by $f$, and it is why we call it \emph{one-shot} decoding.

\begin{figure}[!t]
	\center
	\includegraphics[scale=0.3]{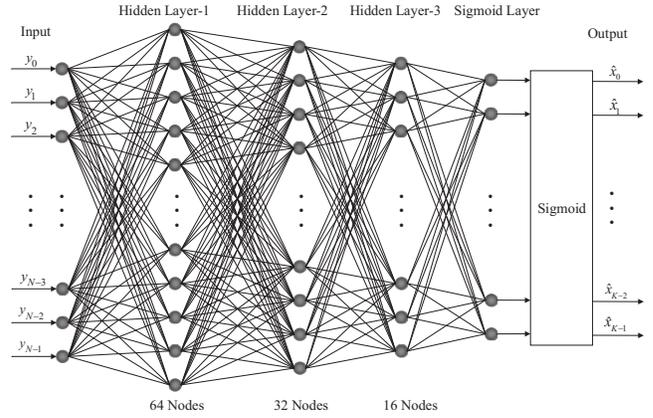}
	\caption{The proposed architecture of multi-layer perceptron.}
	\label{mlp}
\end{figure}

\subsection{Training}
To design the training phase of neural network, which greatly influences the decoding performance of NND, we need solve two problems. First, how to generate training samples. Second, which type of loss function should be chosen.

\subsubsection{Generating training samples}

To train the network, we need both received vector $\textbf{y}$ and true information bits $\textbf{x}$. As such, the general sample generating process can be described like this: The information bits $\textbf{x}$ is randomly picked from $\mathcal{X}$, and then the received vector $\textbf{y}$ can be obtained by performing channel encoding, BPSK mapping and simulated channel noise.

However, there are two important factors we have to consider in the design. One is the ratio of codebook set $\mathcal{X}$ during training phase, we denote it as $p$. If we just randomly pick $\textbf{x}$ from entire set $\mathcal{X}$ in the training phase, the new codewords received at testing phase may have been seen during the training phase, thus the process of neural network is more like recording and reading, other than learning. To evaluate the generalization ability of NND which means if the NND is able to estimate the unseen codewords,  we set that the information bits $\textbf{x}$ is randomly picked from $\mathcal{X}_p$ which covers only $p \%$ of the entire set $\mathcal{X}$. The other factor is the signal to noise ratio
(SNR) of training samples, we denote it as $\rho_{t}$. As the SNR of the actual decoding phase is unknown and time-varying, the performance of NND greatly depends on the SNR of training samples. As in \cite{Gruber2017}, we adopt the proposed method for the setting of training SNR and define a new performance metric which is called the normalized validation error (NVE) as follows

\begin{equation}
\textrm{NVE}(\rho_{t}) = \frac{1}{S}\sum_{s = 1}^{S} \frac{\textrm{BER}_{\textrm{NND}}(\rho_{t}, \rho_{v, s})}{\textrm{BER}_{\textrm{MAP}}(\rho_{v, s})},
\label{NVE}
\end{equation}
where $\rho_{v, s}$ denotes the $s$-th SNR in a set of $S$ different validation samples, $\textrm{BER}_{\textrm{NND}}(\rho_{t}, \rho_{v, s})$ is the bit error rate (BER) achieved by a NND trained at $\rho_{t}$ on the data with $\rho_{v,s}$ and $\textrm{BER}_{\textrm{MAP}}(\rho_{v, s})$ represents the BER of MAP decoding at $\rho_{v,s}$.

As such, the NVE measures how good a NND, trained at a particular SNR, is compared to MAP decoding over a range of different SNRs, and it can be inferred that the less NVE indicates the better performance the NND. As \cite{Gruber2017} says, there is always an optimal $\rho_{t}$, which can be explained by the two extreme cases:
\begin{itemize}
	\item $\rho_{t} \rightarrow +\infty$: train without noise, the NND is not trained
	to handle noise.
	\item $\rho_{t} \rightarrow -\infty$: train only with noise, the NN can not learn
	the encoding structure.
\end{itemize}
This clearly indicates an optimum somewhere in between these two cases. For this reason, we train the NND with datasets of different $\rho_{t}$ in our work, and choose the optimal $\rho_{t}$ which results in the least NVE.

\subsubsection{Cost function}

Loss function is to measure the difference between the actual NND output and its expected output, if the actual output is close to the expected output, the loss should be incremented only slightly whereas large errors should result in a very large loss. In our work, we employ the mean squared error (MSE) as the loss function, which is defined as

\begin{equation}
L_{\textrm{MSE}} = \frac{1}{K} \sum_{i = 0}^{K-1}(x_i - \hat{x}_i)^2,
\label{mse}
\end{equation}
where $x_i \in \{0, 1\}$ is the $i$-th target information bit and $\hat{x}_i$ is the $i$-th NND estimator. Notably, we make $\hat{x}_i \in [0, 1]$ by incurring a sigmoid function at the end of NND, and it can be interpreted as the probability that
a ``1"  was transmitted.

Although there are other commonly used cost functions in neural network, our focus is to compare the NND performance with different types of deep neural network, and the influence of cost function for each type of NND is identical,  thus we just choose the MSE which is simple and easy understanding.

\begin{figure}[!t]
	\center
	\includegraphics[scale=0.32]{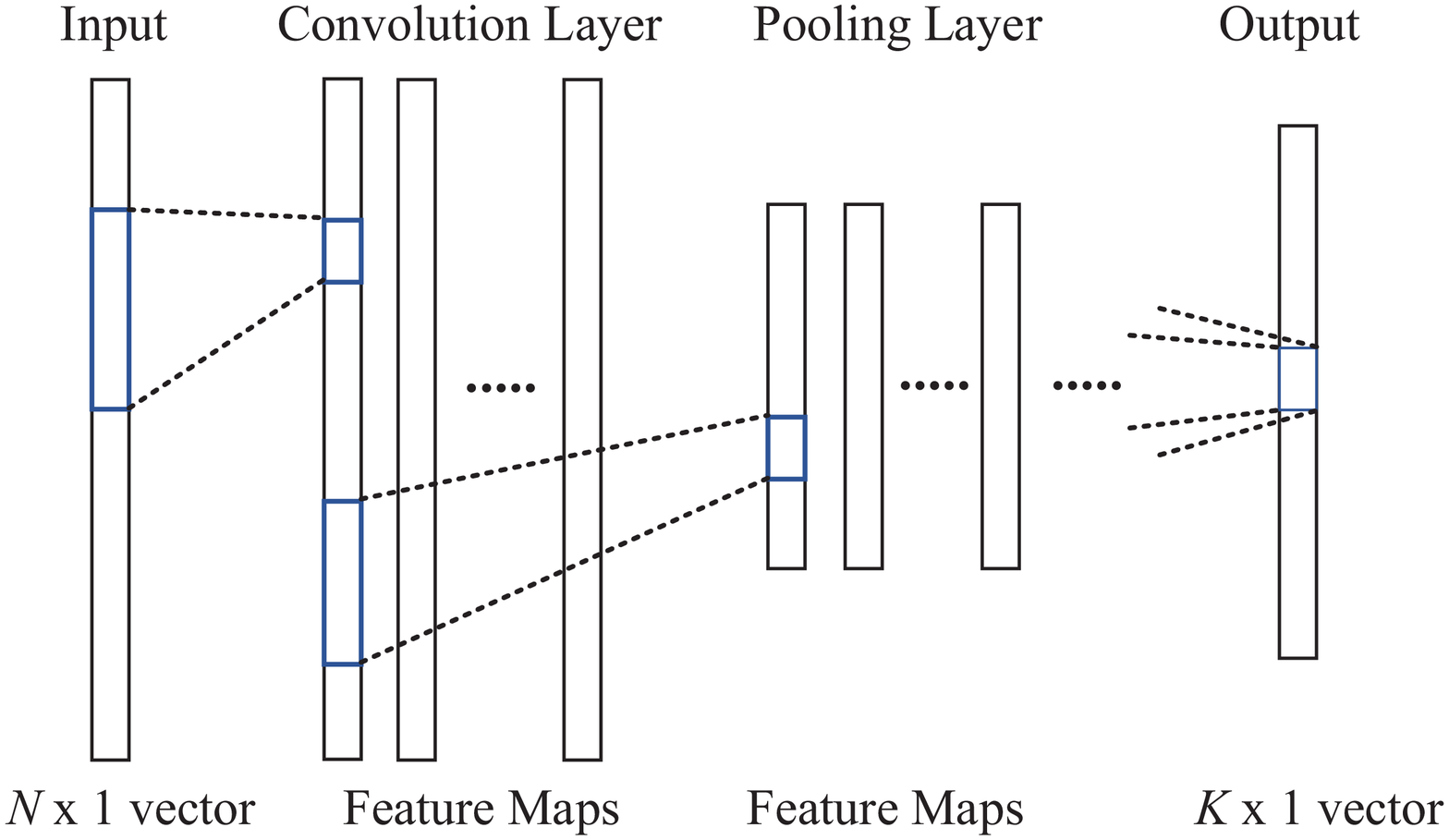}
	\caption{The proposed architecture of convolution neural network.}
	\label{cnn}
\end{figure}

\begin{table}[!hbp]
	\caption{SUMMARY OF TOTAL PARAMETER NUMBER}
	\begin{center}
		\begin{tabular}{ c | c | c | c}
			\hline
			\emph{Number of $N$} & \emph{MLP}  &  \emph{CNN}   & \emph{RNN} \\ \hline
			$N = 8$ & $3136$  & $2072$  & $2052$ \\ \hline
			$N = 16$ & $3712$  & $2456$  & $3076$ \\ \hline
			$N = 32$ & $4864$  & $3992$  & $5124$ \\ \hline
		\end{tabular}		
	\end{center}
	\label{parameter}
\end{table}

\subsection{Design of Deep Neural Network}

In this section, our specific designs for MLP, CNN and RNN are described. It is well known that if the parameters of neural network (weights and biases) are very large, the expression and learning ability of neural network usually will be very strong. Considering that our aim is to compare the learning ability of different deep neural networks, we should keep the total parameter number of each neural network approximately the same, to avoid that the performance difference of deep neural networks comes from the difference of parameter number. As such, we construct a relative simple and general structure for each neural network, and keep the magnitude of parameters as $10^3$, which is shown in Table \ref{parameter} .

MLP is a class of feedforward artificial neural network with fully connection between layers, which consists of at least three layers of nodes. Each node of MLP is a neuron that uses a nonlinear activation function which gives it learning ability. It is shown in \cite{Hornik1989} that nonlinear activation functions can theoretically approximate any continuous function on a bounded region arbitrarily closely if the number of neurons is large enough. In this paper, the  proposed architecture of MLP is described in Fig. \ref{mlp}. We employ three hidden layers with 64, 32, 16 in MLP, and the nodes of input layer and output layer is $N$ and $K$ with no doubt.

CNN is also a class of feedforward artificial neural networks which has been successfully applied to analyzing visual imagery. The hidden layers of CNN are either convolutional or pooling, which emulated the response of an individual neuron to visual stimuli \cite{LuCun1998}, and the convolution operation significantly reduces the number of parameters, allowing the network to be deeper with fewer parameters. As such, the excellent performance of CNN for the image characteristic extraction motivates us to combine CNN with NND in this work. Considering that CNN is usually used for image, some modifications are needed when applying CNN in NND. As Fig. \ref{cnn} shows, we modify the each layer's input of CNN as a 1-D vector instead of a 2-D image. Also, we employ a general structure of CNN without some high-level tricks like batch normalization presented in \cite{Ioffe2015}, the detailed parameter setting is listed in Table \ref{table_cnn}.

\begin{table}[!t]
	\caption{PARAMETERS of CNN}
	\begin{center}
		\begin{tabular}{ c | c | c }
			\hline
			\emph{Type of layer} & \emph{Kernel size / stride (or Annotation)}  &  \emph{Input size}\\ \hline
			Convolution & $3 \times 1 ~/~ 1$ &  $N \times 1 \times 1$ \\ \hline
			Pooling & $2 \times 1 ~/~ 2$  &  $N \times 1 \times 8$ \\ \hline
			Convolution & $3 \times 1 ~/~ 1$ &  $N/2 \times 1 \times 8$ \\ \hline
			Pooling & $2 \times 1 ~/~ 2$  &  $N/2 \times 1 \times 16$ \\ \hline
			Convolution & $3 \times 1 ~/~ 1$ &  $N/4 \times 1 \times 16$ \\ \hline
			Pooling & $2 \times 1 ~/~ 2$  &  $N/4 \times 1 \times 32$ \\ \hline
			Convolution & $N/8 \times 1 ~/~ 1$ &  $N/8 \times 1 \times 32$ \\ \hline
			Squeeze & Reduce the dimension  &  $1 \times 1 \times K$ \\	\hline	
			Sigmoid &  Output the classification probability &  $1 \times K$ \\		
			\hline
		\end{tabular}		
	\end{center}
	\label{table_cnn}
\end{table}

\begin{figure}[!t] \centering
	\subfigure[Basic LSTM cell] { \label{fig:a}
		\includegraphics[scale = 0.2]{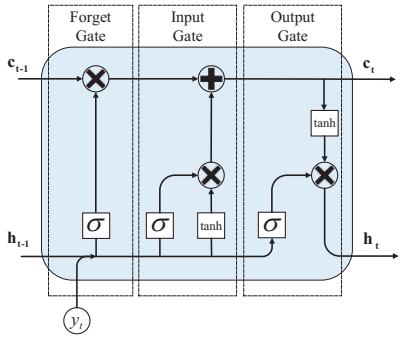}
	}
	\subfigure[The flow diagram of LSTM.] { \label{fig:b}
		\includegraphics[scale = 0.3]{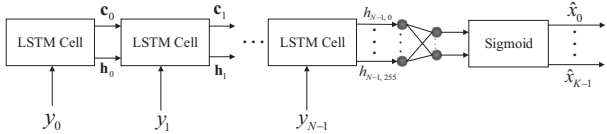}
	}
	\caption{The proposed architecture of long short-term memory (LSTM). }
	\label{rnn}
\end{figure}

RNN is a class of artificial neural network where connections between units form a directed cycle. This allows it to exhibit dynamic temporal behavior. Unlike feedforward neural networks, RNN can use their internal memory to process arbitrary sequences of inputs. This makes them applicable to tasks such as unsegmented, connected handwriting recognition or speech recognition \cite{Shelke2010}. Inspired by the remarkable performance of RNN on the time series task, we hope it can also achieve a good performance in NND. Notably, general RNN suffers a serious vanishing gradient problem as described in \cite{Gradient2009}, people usually adopt long short-term memory (LSTM) in practice. As presented in \cite{Hochreiter1997}, LSTM contains three gates, called forget gate, input gate and output gate respectively, to control the information flow, which can prevent backpropagated errors from vanishing or exploding. As such, we take LSTM as the representative of RNN, and the proposed architecture is described in Fig. \ref{rnn}. We set the output dimension of LSTM cell is 256, i.e., $h$ and $c$ are $256 \times 1$ vectors, and only one LSTM cell is employed in each time step.

\begin{table}[!hbp]
	\caption{HYPERPARAMETER SETTING}
	\begin{center}
		\begin{tabular}{ c | c }
			\hline
			Size of training samples  & $10^6$  \\ \hline	
			Size of testing samples  & $10^5$  \\ \hline			
			Mini-batch size  & $128$  \\ \hline	
			Dropout probability  & $0.1$  \\ \hline	
			Initialization method  & Xavier initialization  \\ \hline	
			Optimization method  & Adam optimization  \\ \hline			
		\end{tabular}		
	\end{center}
	\label{hyperparameter}
\end{table}

\section{performance evaluation}  \label{performance evaluation}

In this section, the performances of NND with MLP, CNN and RNN for different length of $N$ are compared. Throughout all experiments, we use a polar code of rate $1/2$ and set the codeword length $N$ as $8, 16, 32$. The training ratio of codebook $p$ is set as $40 \%, 60 \%, 80 \%$ and $100 \%$. We take $12$ different SNR points from $-2$ dB to $20$ dB as training SNR $\rho_t$, and as mentioned before,  the best $\rho_t$, which results in the least NVE, will be chosen for testing phase. For the setting of general parameters in neural network, which are called hyperparameters \cite{Bergstra2011}, we choose a relatively reasonable and satisfying set as shown in Table \ref{hyperparameter} after a lot of trials. Notably, we use Tensorflow as our experimental platform, and the source code\footnote{https://github.com/levylv/deep-neural-network-decoder} is available for reproducible research.

To better evaluate the learning ability of NND, we divide the simulation into two parts. First part investigates the performance of NND that are trained without noise, which only reflects the learning ability for encoding structure. Second part investigates the performance of NND that are trained with noise, which involves the simultaneous learning for encoding structure and noise characteristic. Additionally, the system performance is measured by the bit error rate (BER) based on the testing samples, and the performance of MAP decoding is compared as a benchmark.

\begin{figure} \centering
	\subfigure[MLP] { \label{fig_no_mlp_8}
		\includegraphics[scale = 0.3]{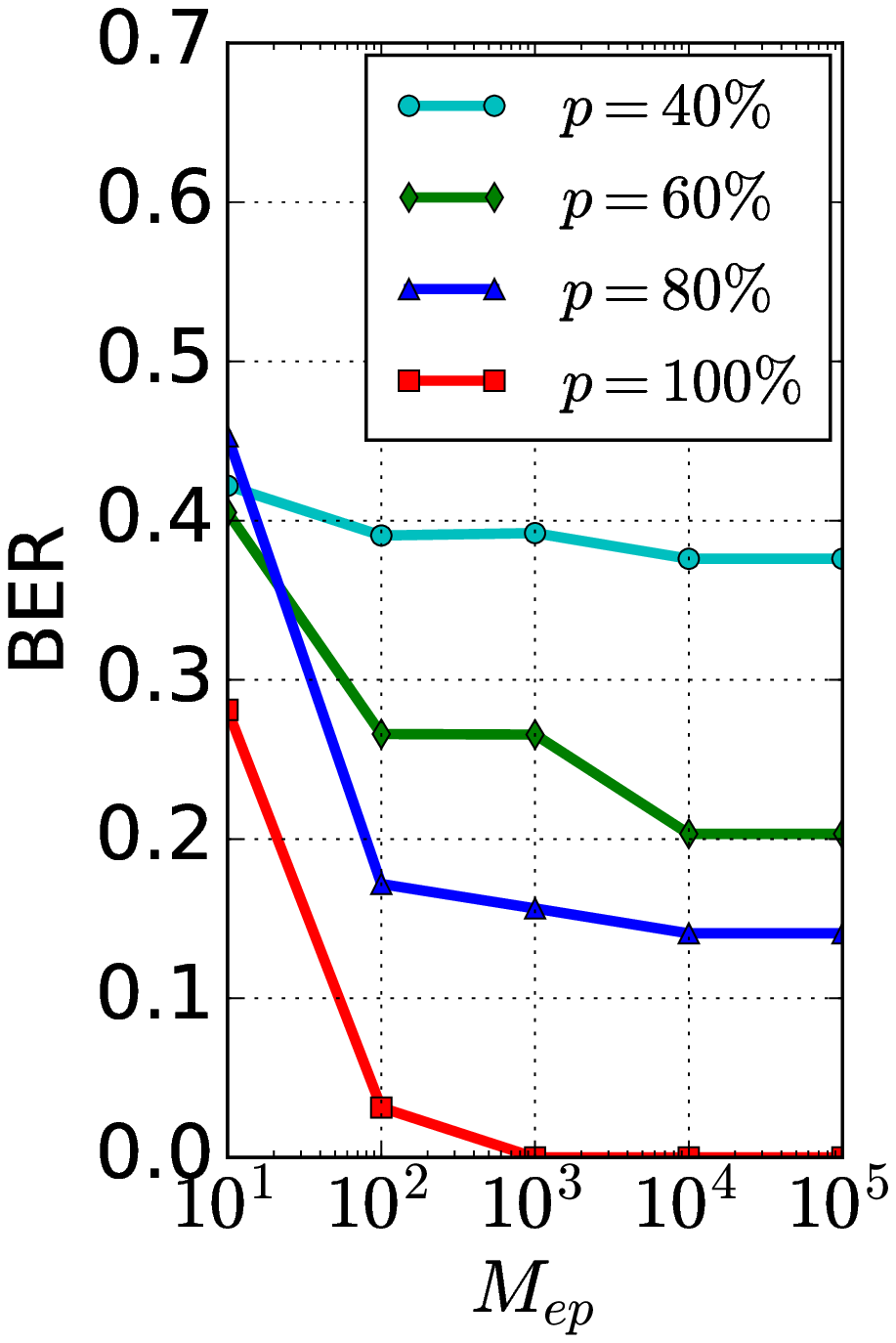}
	}
	\hspace{-0.65cm}
	\subfigure[CNN] { \label{fig_no_cnn_8}
		\includegraphics[scale = 0.3]{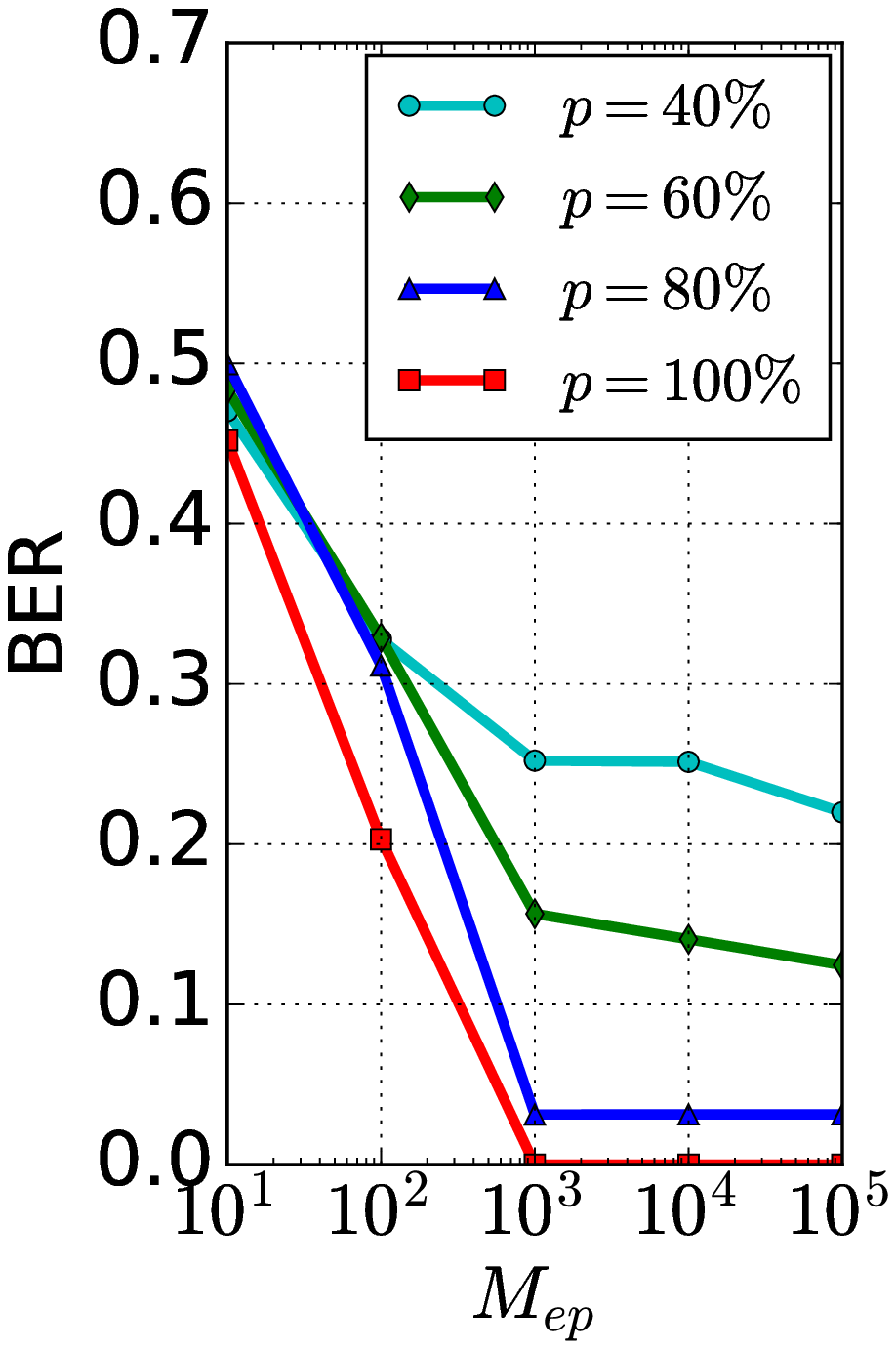}
	}
	\hspace{-0.65cm}
	\subfigure[RNN] { \label{fig_no_rnn_8}
		\includegraphics[scale = 0.3]{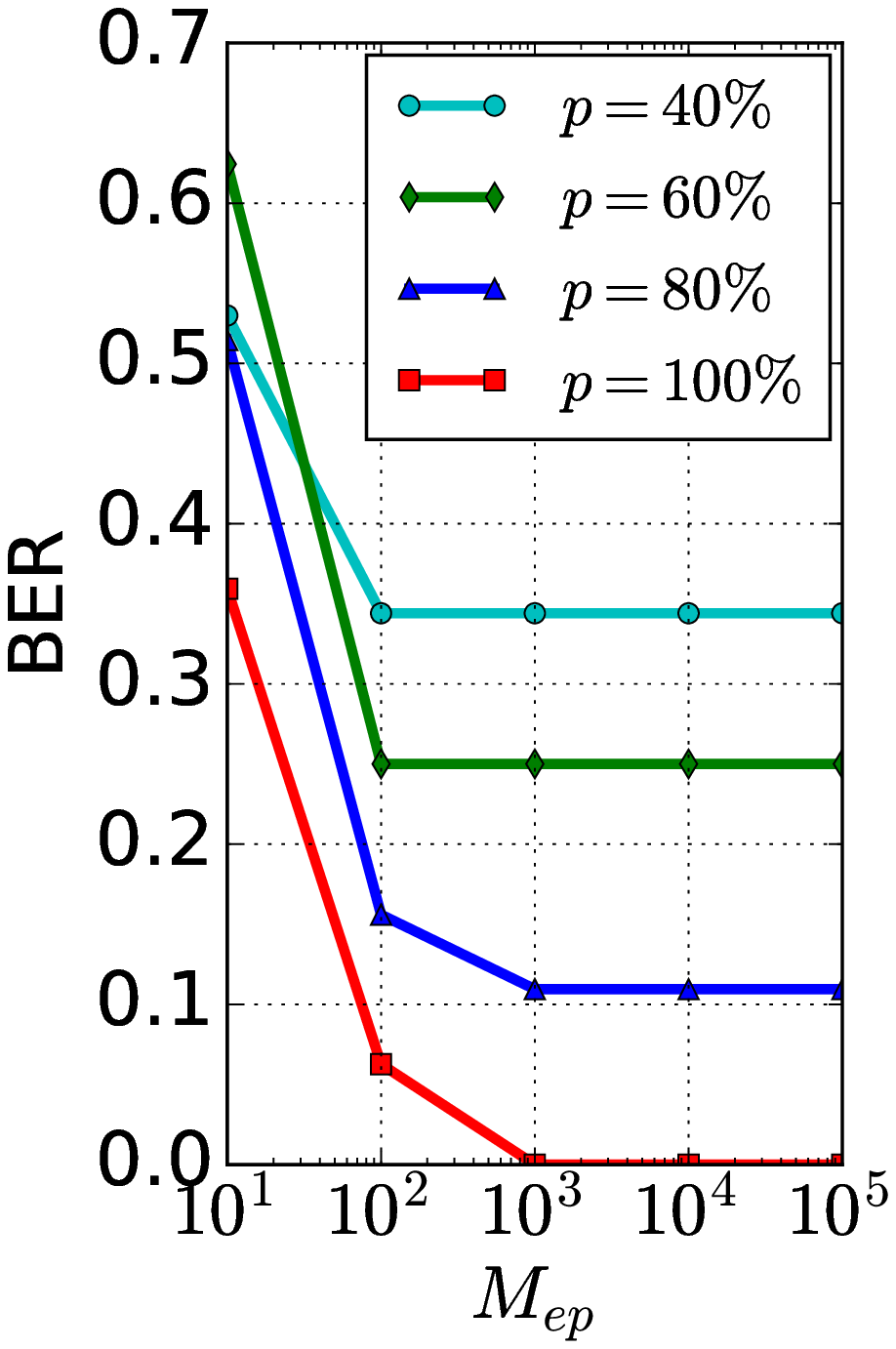}
	}
	\caption{The BER achieved by MLP, CNN and RNN without noise versus the number of training epoch $M_{ep} $ for $N = 8$ with training ratio $p = 40\%, 60\%, 80\%$ and $100\%$.}
	\label{fig_no_8}
\end{figure}

\begin{figure} \centering
	\subfigure[MLP] { \label{fig_no_mlp_16}
		\includegraphics[scale = 0.3]{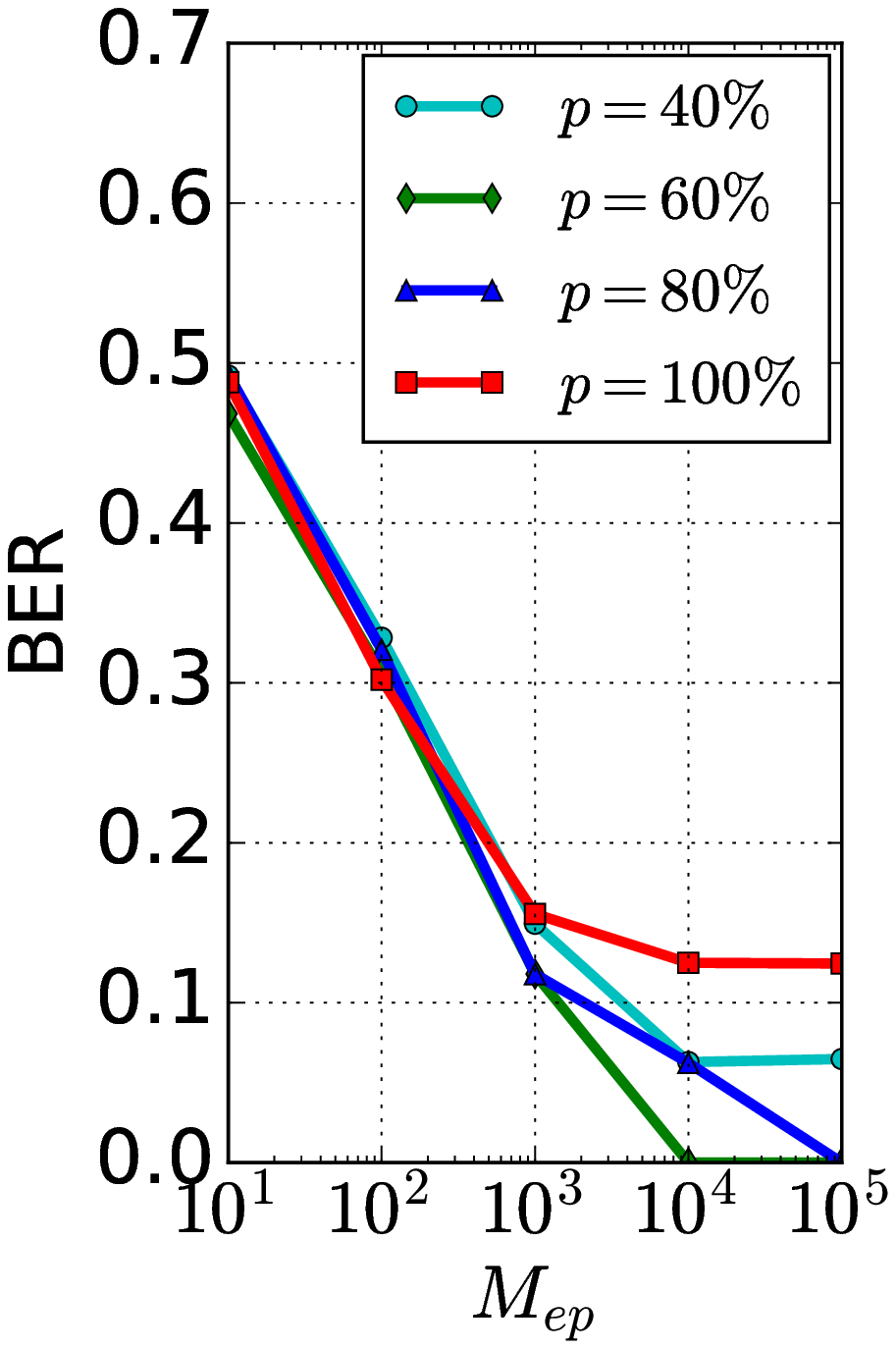}
	}
	\hspace{-0.65cm}
	\subfigure[CNN] { \label{fig_no_cnn_16}
		\includegraphics[scale = 0.3]{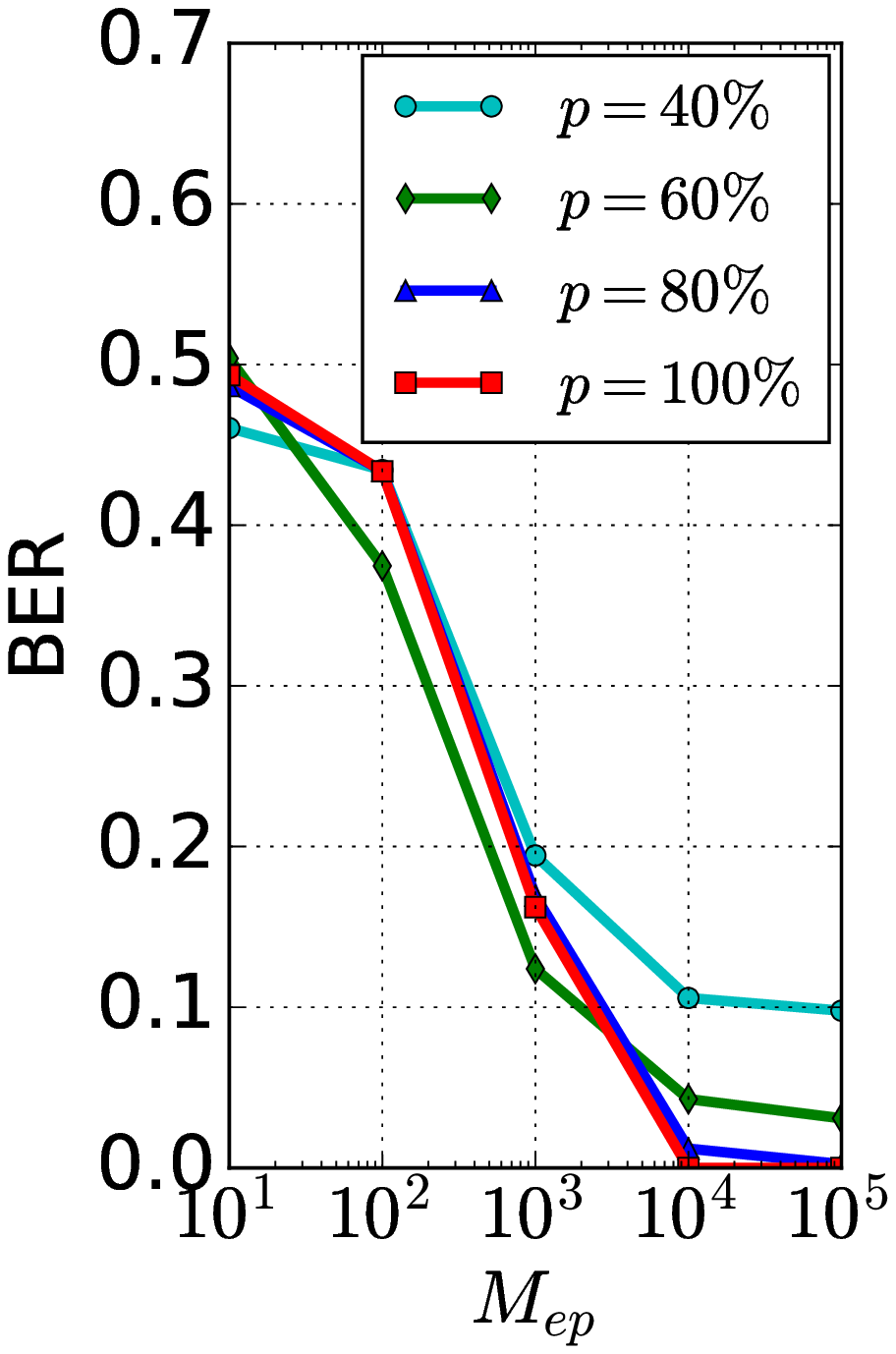}
	}
	\hspace{-0.65cm}
	\subfigure[RNN] { \label{fig_no_rnn_16}
		\includegraphics[scale = 0.3]{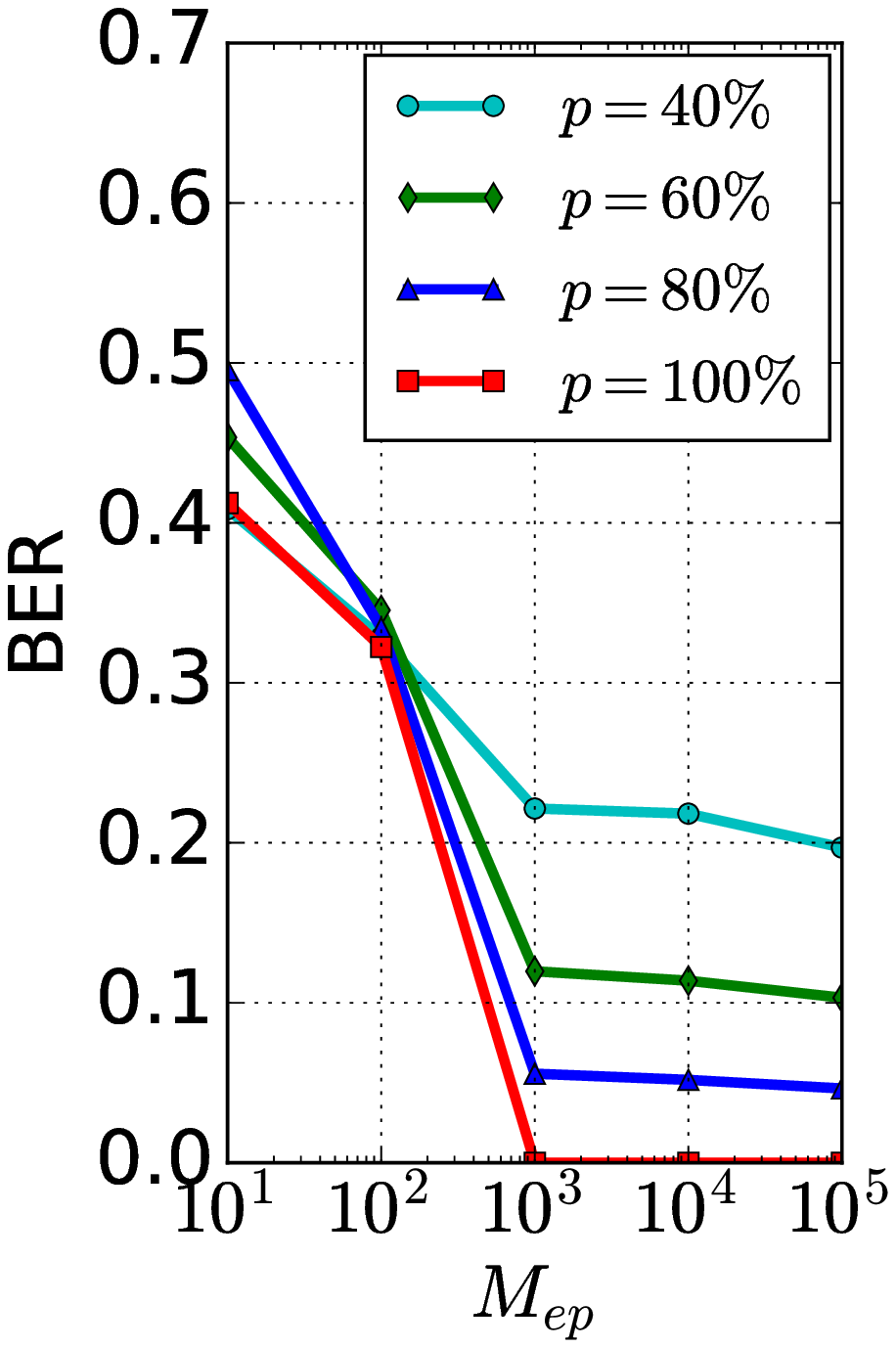}
	}
	\caption{The BER achieved by MLP, CNN and RNN without noise versus the number of training epoch $M_{ep} $ for $N = 16$ with training ratio $p = 40\%, 60\%, 80\%$ and $100\%$.}
	\label{fig_no_16}
\end{figure}

\subsection{Learning without Noise}

Fig. \ref{fig_no_8} investigates the BER achieved by MLP, CNN and RNN without noise as a function of the number of training epochs ranging from $M_{ep} = 10^1, \dots, 10^5$ for $N = 8$, with respect to different training ratio of codebook. From Fig. \ref{fig_no_8}, we can see that the BER decreases gradually with the number of training epoch increasing, and it finally reaches a steady value, which represents the convergence of deep neural network. For all three deep neural networks, it can be observed that the lower $p$ leads to higher BER, and only when $p = 100\%$ the BER drops to $0$, which denotes the neural networks have learned the complete encoding structure. The phenomenon can be explained as the overfitting of neural network, i.e., the neural network can fit the relationship between input and output very well for each value of $p$ even if it is not $100\%$, thus it leads to a bad generalization performance, which is based on the entire codebook.

The similar case but when $N = 16$ is studied in Fig. \ref{fig_no_16}. For MLP, we find that when $p = 100\%$, the performance is getting worse compared with the case of $N = 8$, the reason is the size of codebook exponentially increase with the growth of $N$, thus it induces a more complex relationship between the input and output of NND, which goes beyond the learning ability of current neural network and make it tend to be underfitting. However, it is satisfied to see that when $p = 60\%$ and $80\%$, the BER drops to $0$, which represents the MLP is able to generalize from a fraction of codebook to the entire codebook set. The only difference of CNN when compared with MLP is that when BER drops to $0$, the $p$ is $80\%$ and $100\%$, it indicates the learning ability of CNN is a little bit stronger than MLP. Notably, the RNN is still in the state of overfitting since its performance remains the same as the case when $N = 8$.

\begin{figure} \centering
	\subfigure[MLP] { \label{fig_no_mlp_32}
		\includegraphics[scale = 0.3]{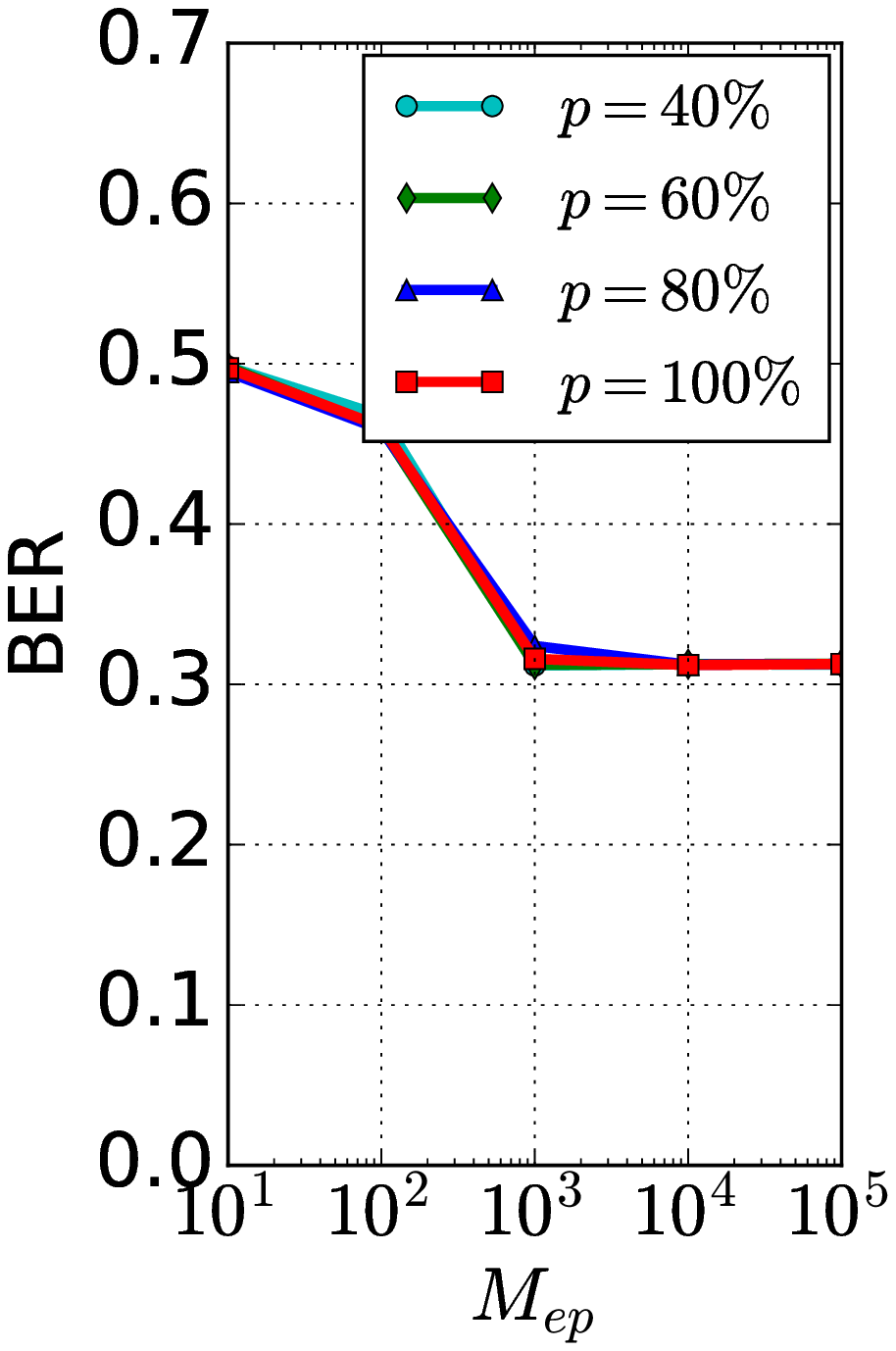}
	}
	\hspace{-0.65cm}
	\subfigure[CNN] { \label{fig_no_cnn_32}
		\includegraphics[scale = 0.3]{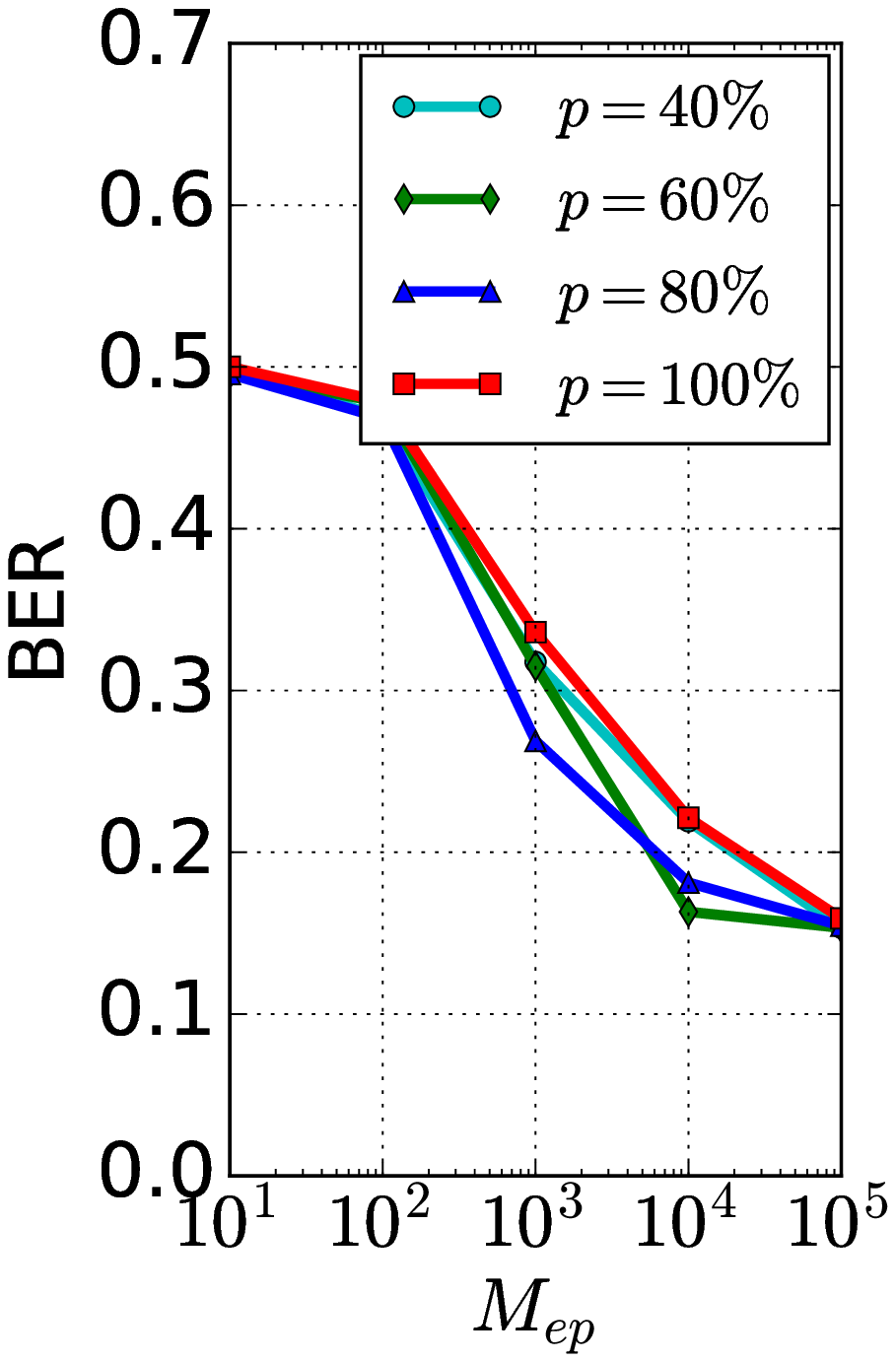}
	}
	\hspace{-0.65cm}
	\subfigure[RNN] { \label{fig_no_rnn_32}
		\includegraphics[scale = 0.3]{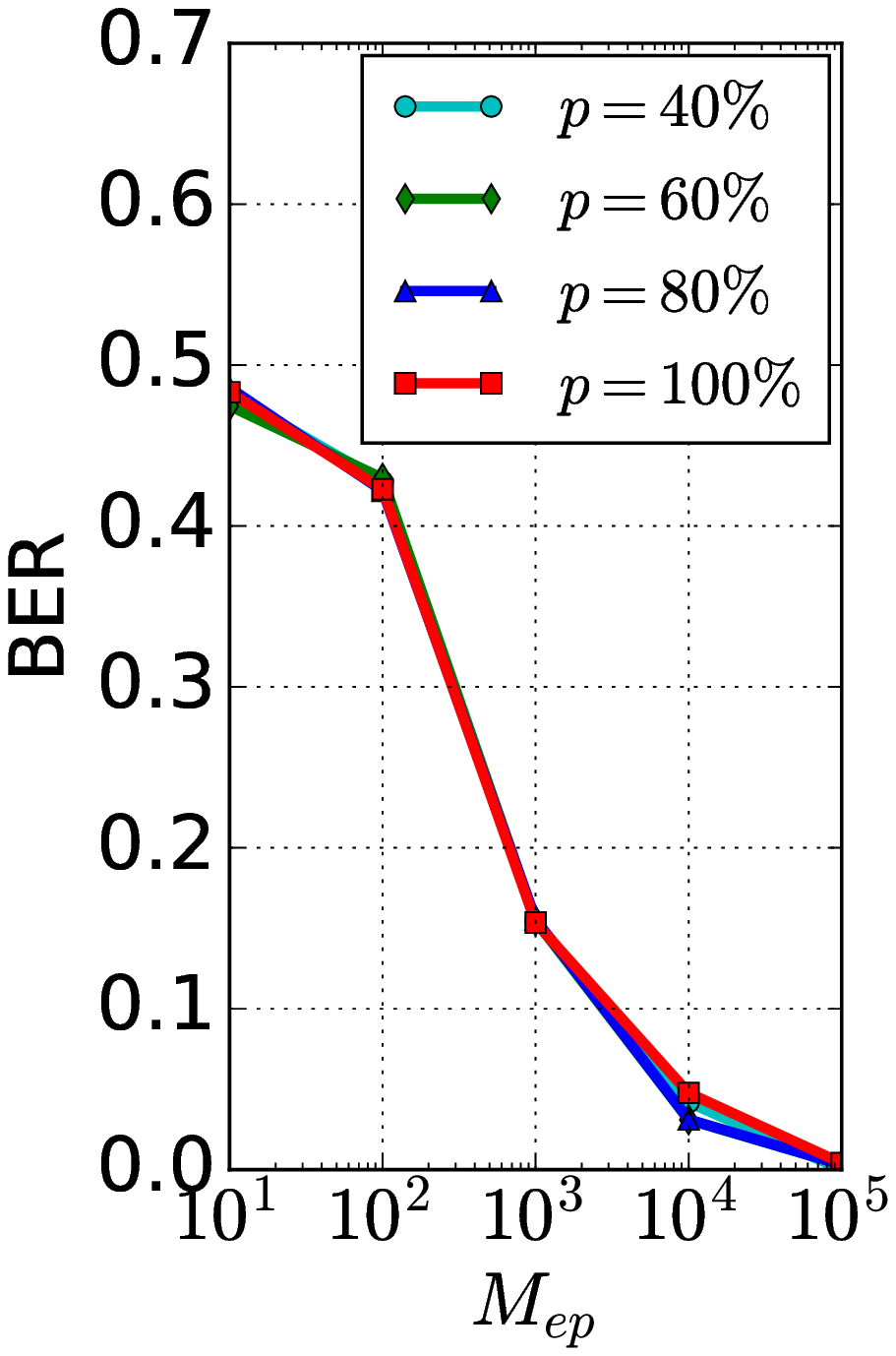}
	}
	\caption{The BER achieved by MLP, CNN and RNN without noise versus the number of training epoch $M_{ep} $ for $N = 32$ with training ratio $p = 40\%, 60\%, 80\%$ and $100\%$.}
	\label{fig_no_32}
\end{figure}

Similarly, Fig. \ref{fig_no_32} is for the case when $N = 32$. Obviously, it can be inferred that both of MLP and CNN are in state of underfitting. However, RNN can achieve a very good performance because the BER drops to $0$ with each value of $p$, it proves that RNN is better than MLP and CNN for NND.

\subsection{Learning with Noise}
In this part, we research the case with noise which involves the simultaneous learning for encoding structure and noise characteristic, and the BER performances versus the testing SNR $E_b / N_0$ for different codeword length $N$ are investigated.

\begin{figure} \centering
	\subfigure[MLP] { \label{fig_mlp_8}
		\includegraphics[scale = 0.29]{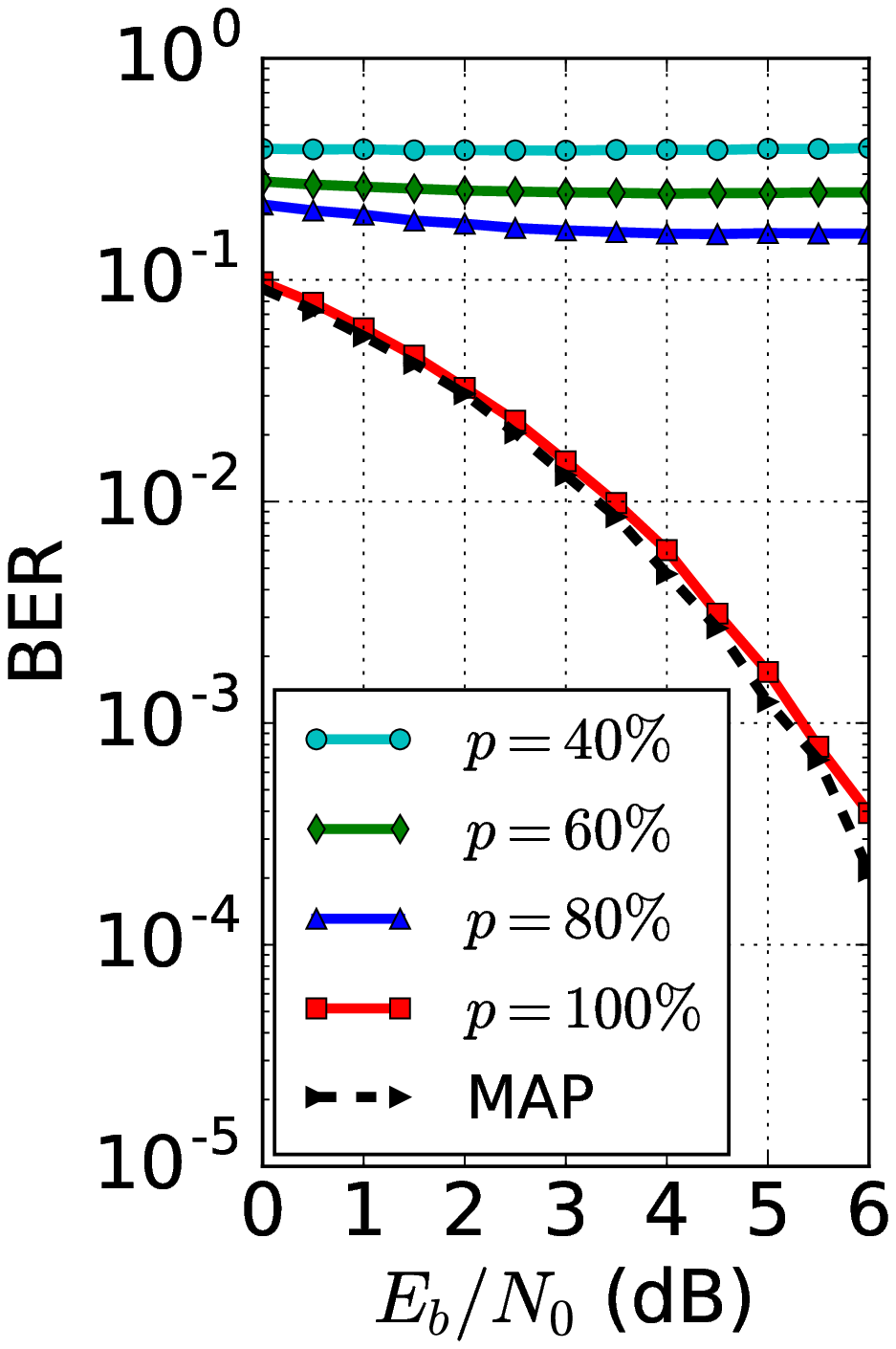}
	}
	\hspace{-0.65cm}
	\subfigure[CNN] { \label{fig_cnn_8}
		\includegraphics[scale = 0.29]{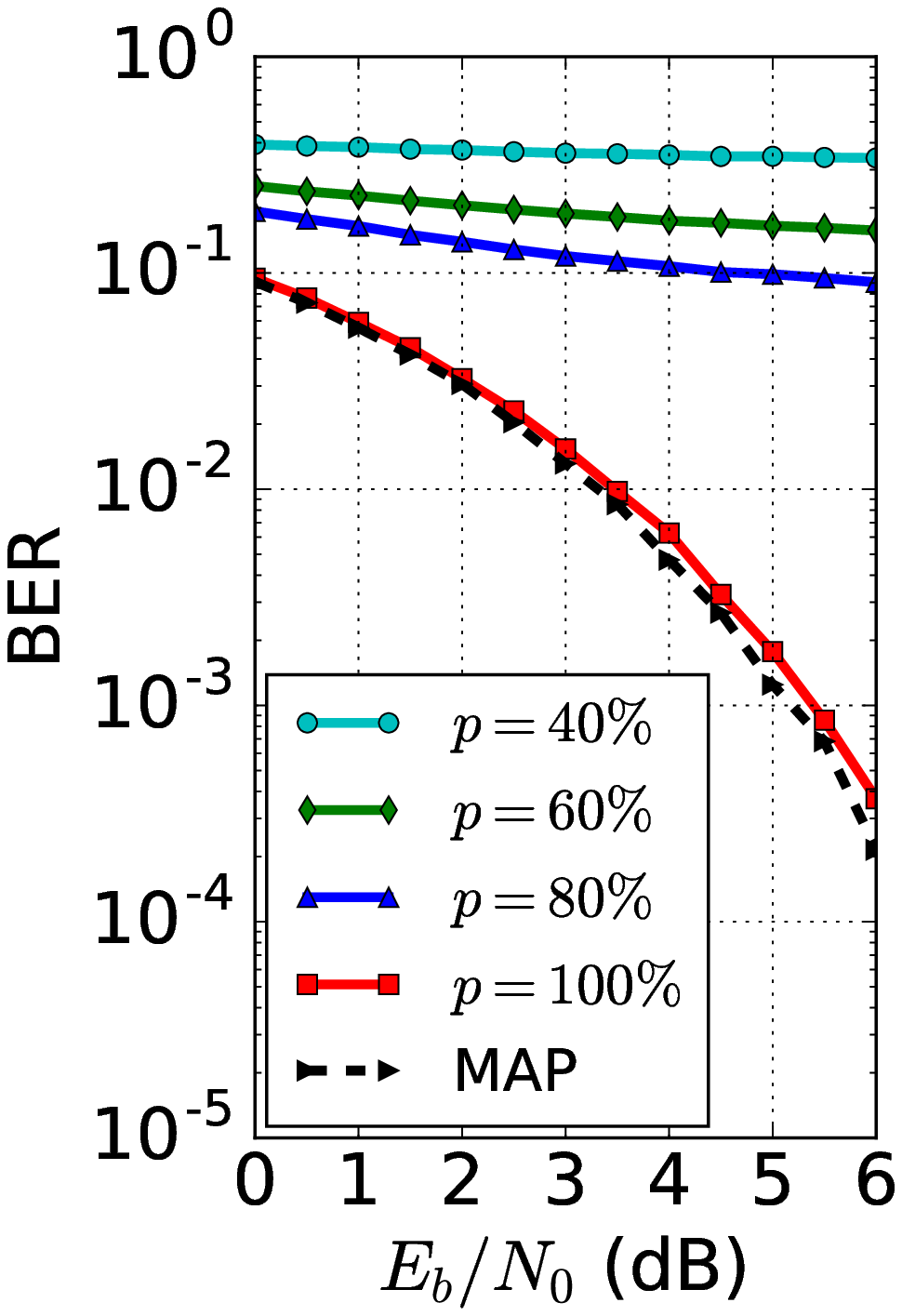}
	}
	\hspace{-0.65cm}
	\subfigure[RNN] { \label{fig_rnn_8}
		\includegraphics[scale = 0.29]{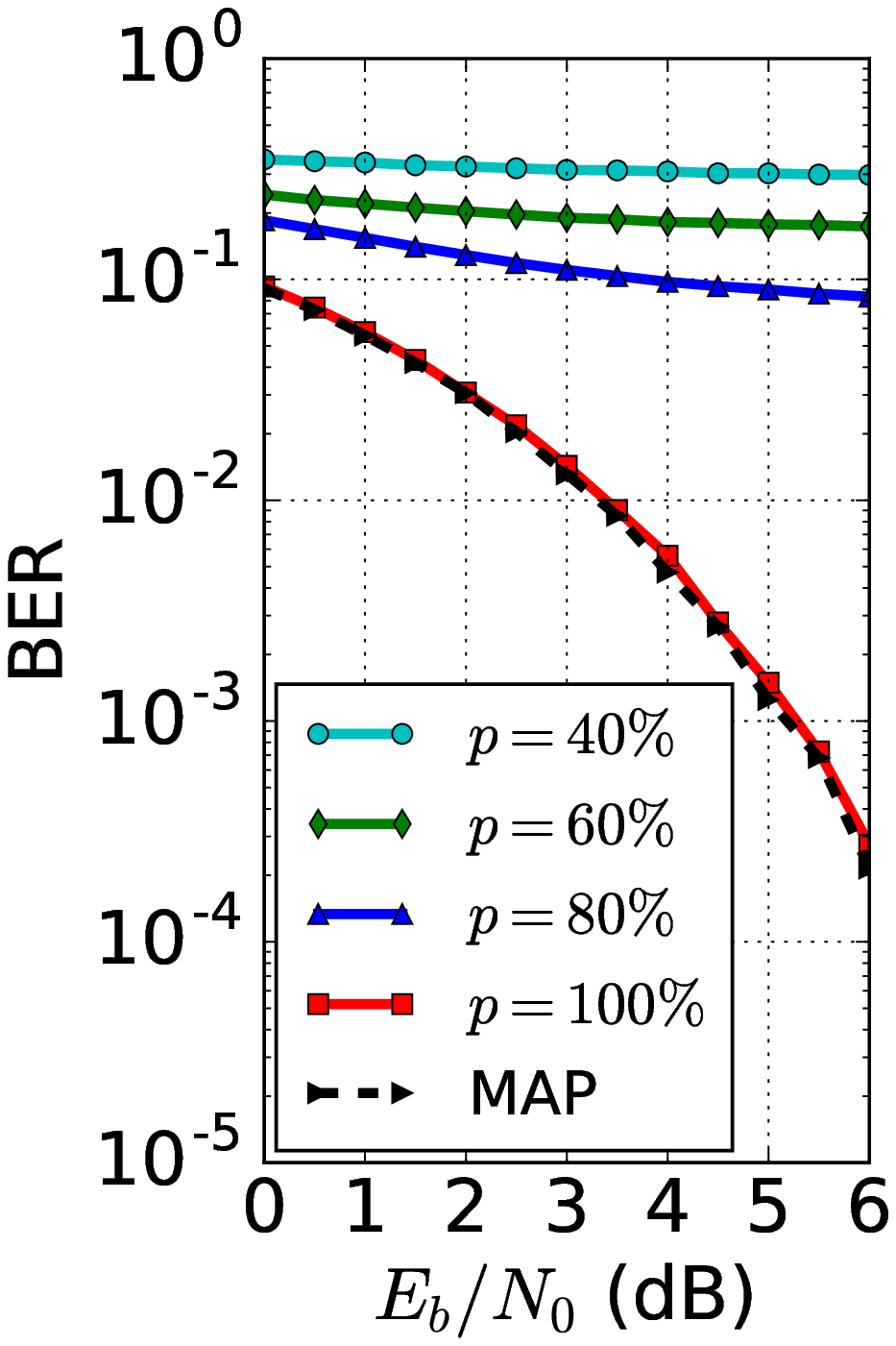}
	}
	\caption{The BER achieved by MLP, CNN and RNN with noise versus the testing SNR $E_b / N_0$ for $N = 8$ with training ratio $p = 40\%, 60\%, 80\%$ and $100\%$ and $M_{ep} = 10^5$.}
	\label{fig_8}
\end{figure}

\begin{figure} \centering
	\subfigure[MLP] { \label{fig_mlp_16}
		\includegraphics[scale = 0.29]{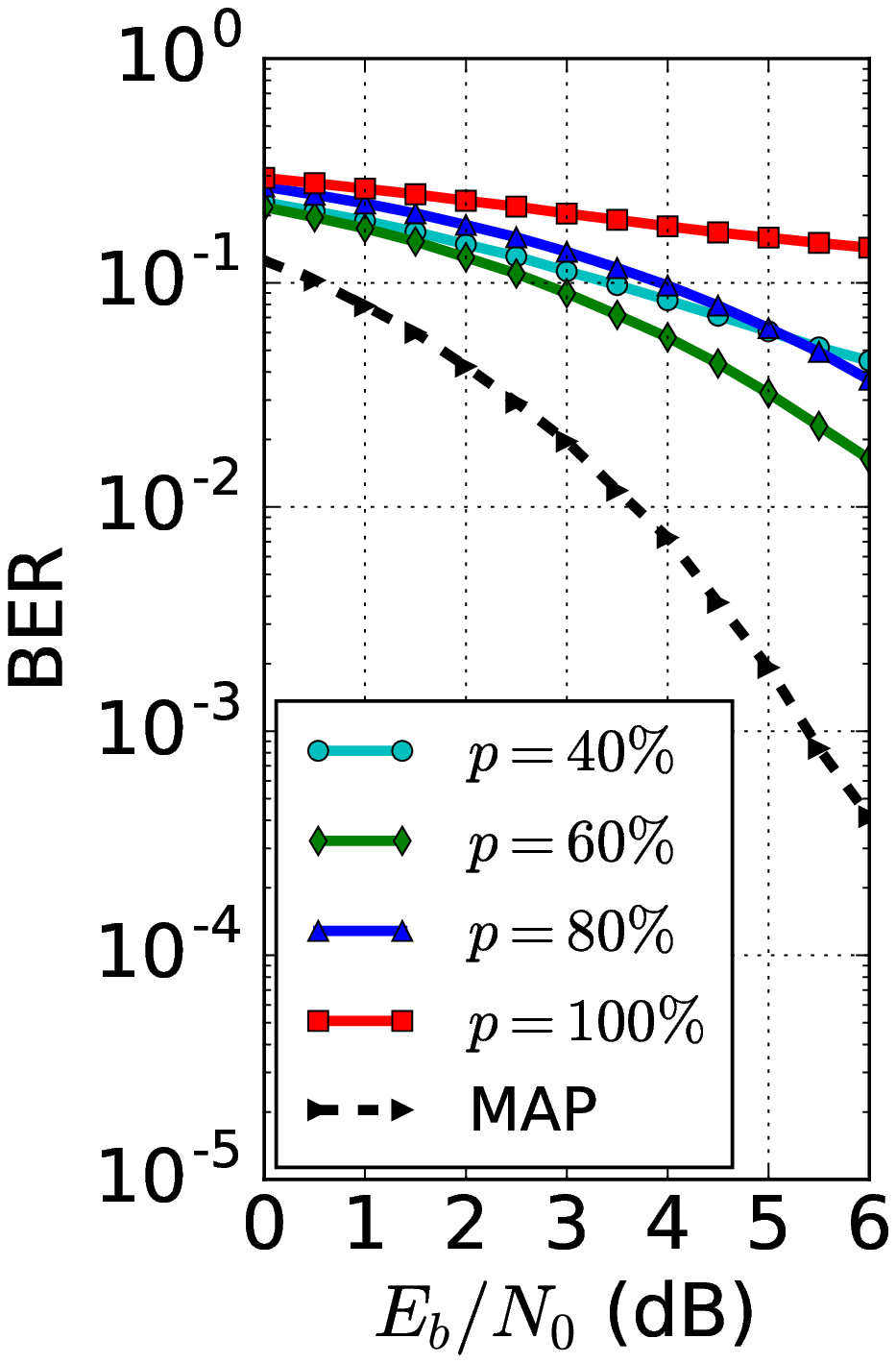}
	}
	\hspace{-0.65cm}
	\subfigure[CNN] { \label{fig_cnn_16}
		\includegraphics[scale = 0.29]{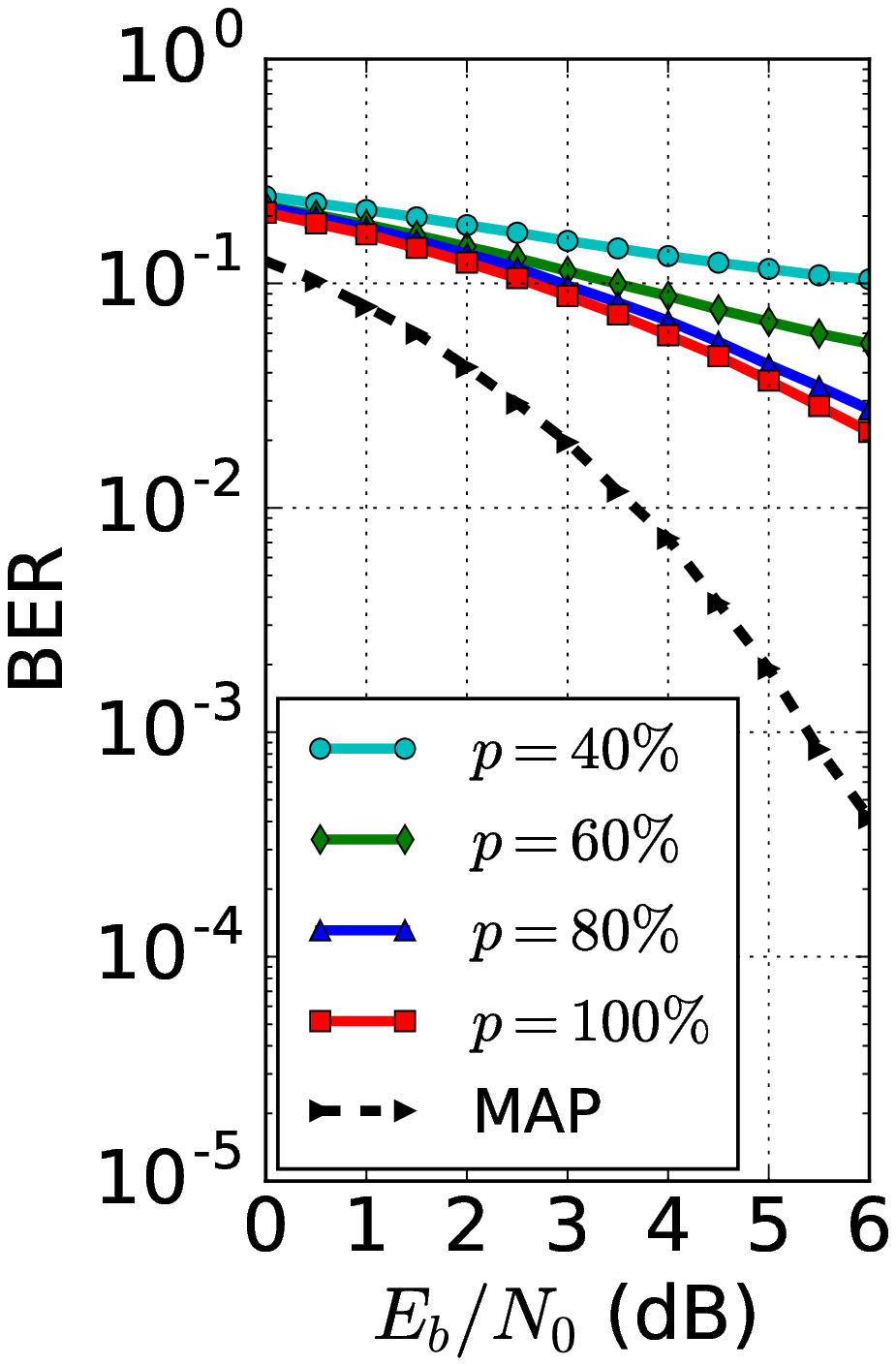}
	}
	\hspace{-0.65cm}
	\subfigure[RNN] { \label{fig_rnn_16}
		\includegraphics[scale = 0.29]{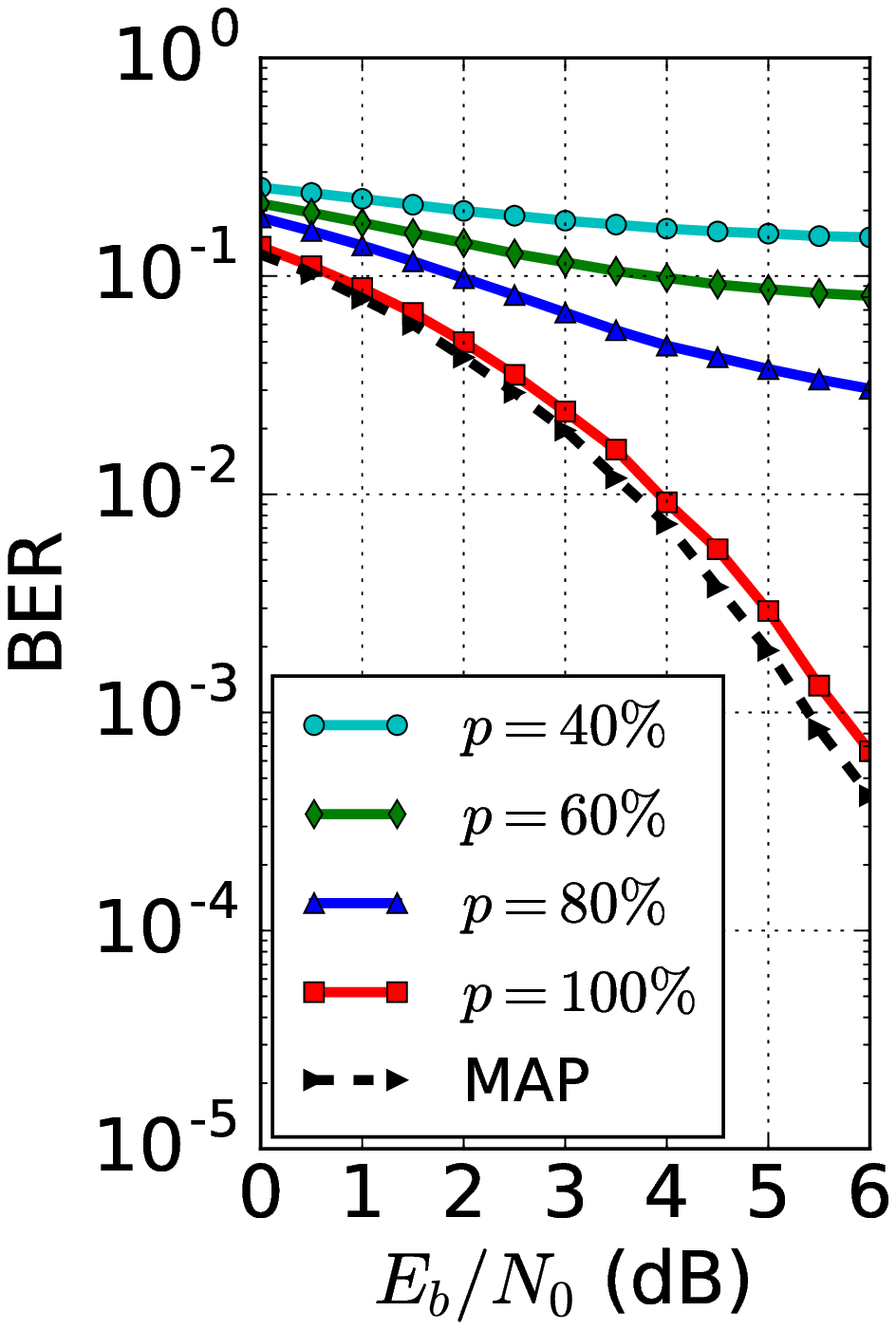}
	}
	\caption{The BER achieved by MLP, CNN and RNN with noise versus the testing SNR $E_b / N_0$ for $N = 16$ with training ratio $p = 40\%, 60\%, 80\%$ and $100\%$ and $M_{ep} = 10^5$.}
	\label{fig_16}
\end{figure}

\begin{figure} \centering
	\subfigure[MLP] { \label{fig_mlp_32}
		\includegraphics[scale = 0.28]{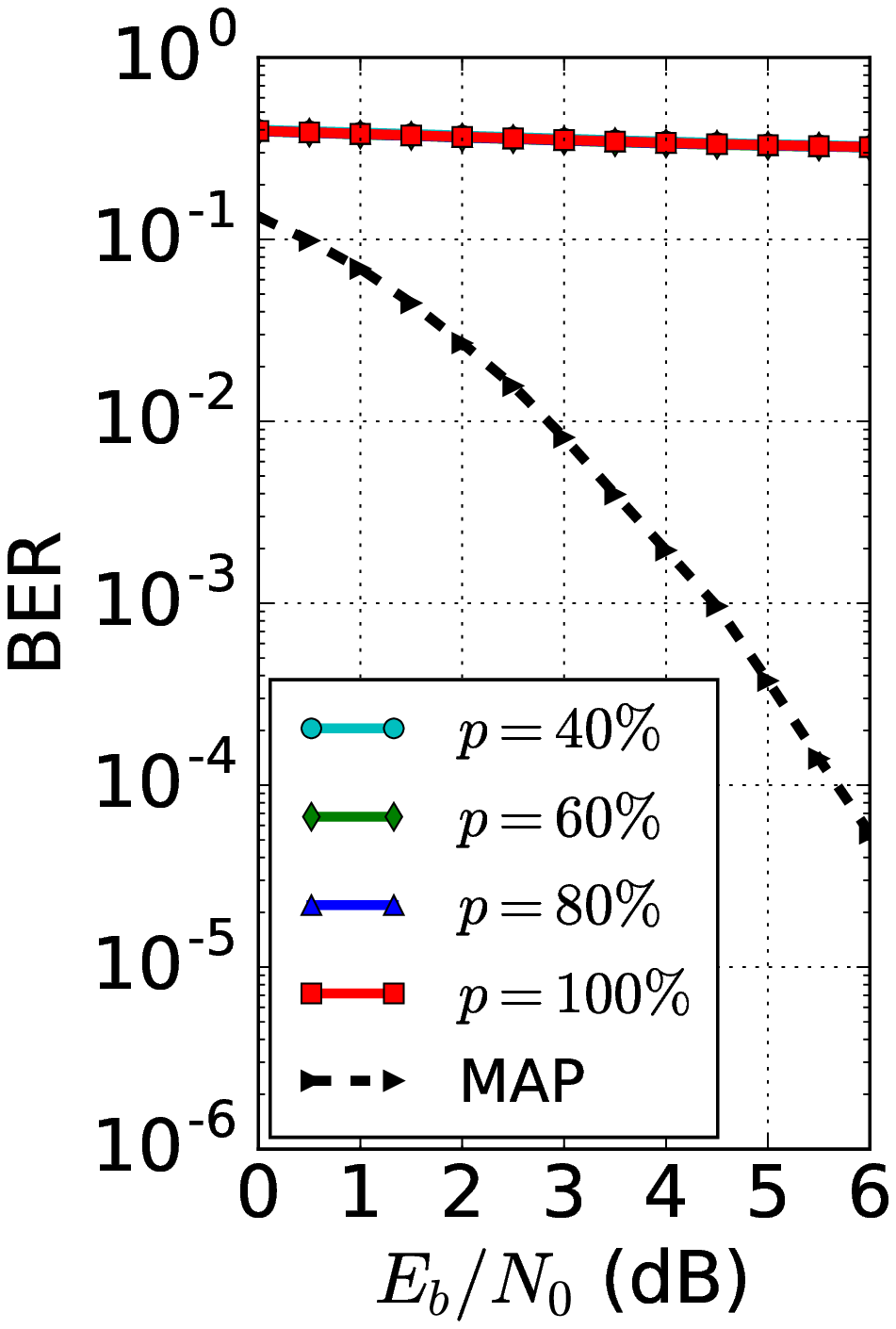}
	}
	\hspace{-0.65cm}
	\subfigure[CNN] { \label{fig_cnn_32}
		\includegraphics[scale = 0.28]{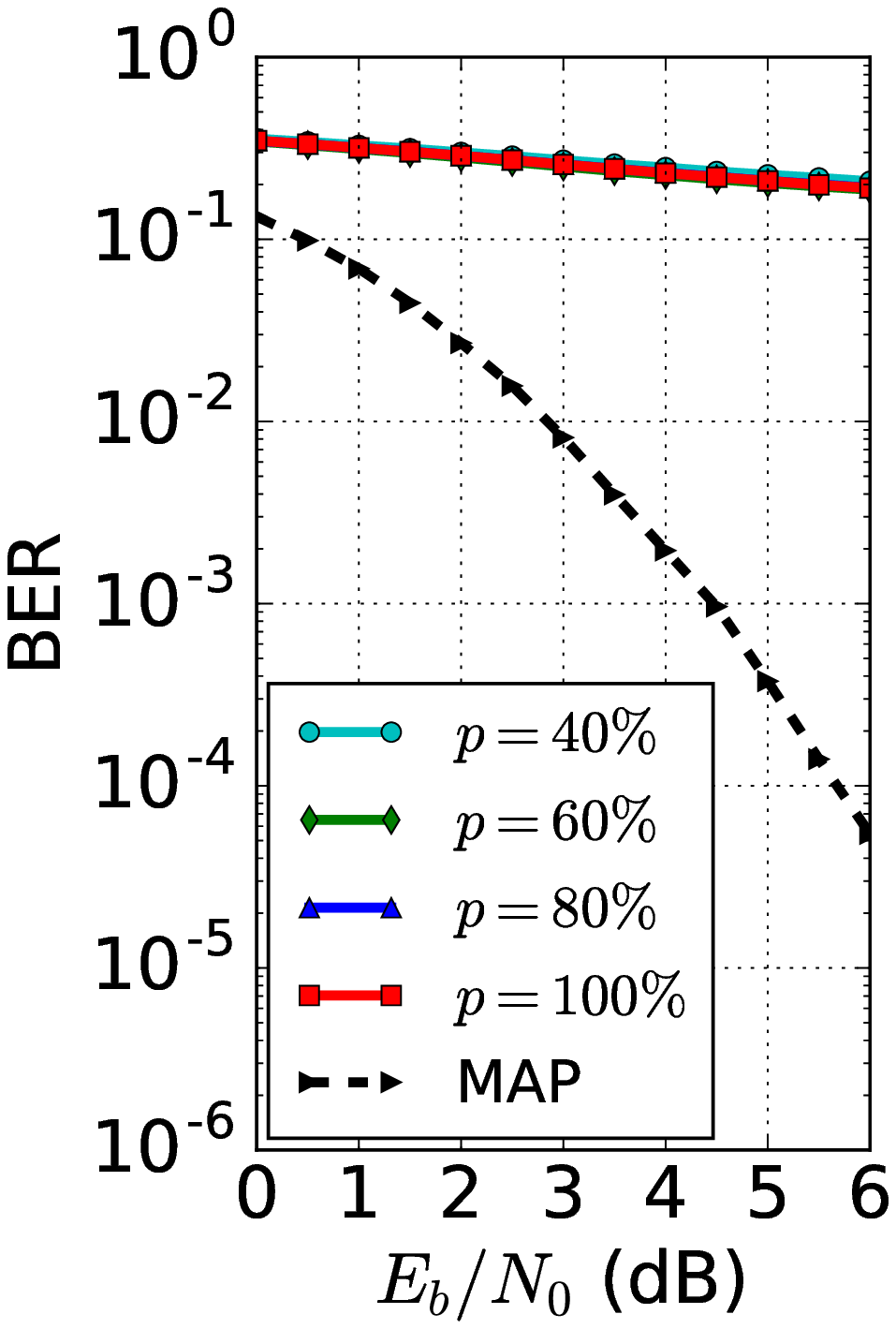}
	}
	\hspace{-0.65cm}
	\subfigure[RNN] { \label{fig_rnn_32}
		\includegraphics[scale = 0.28]{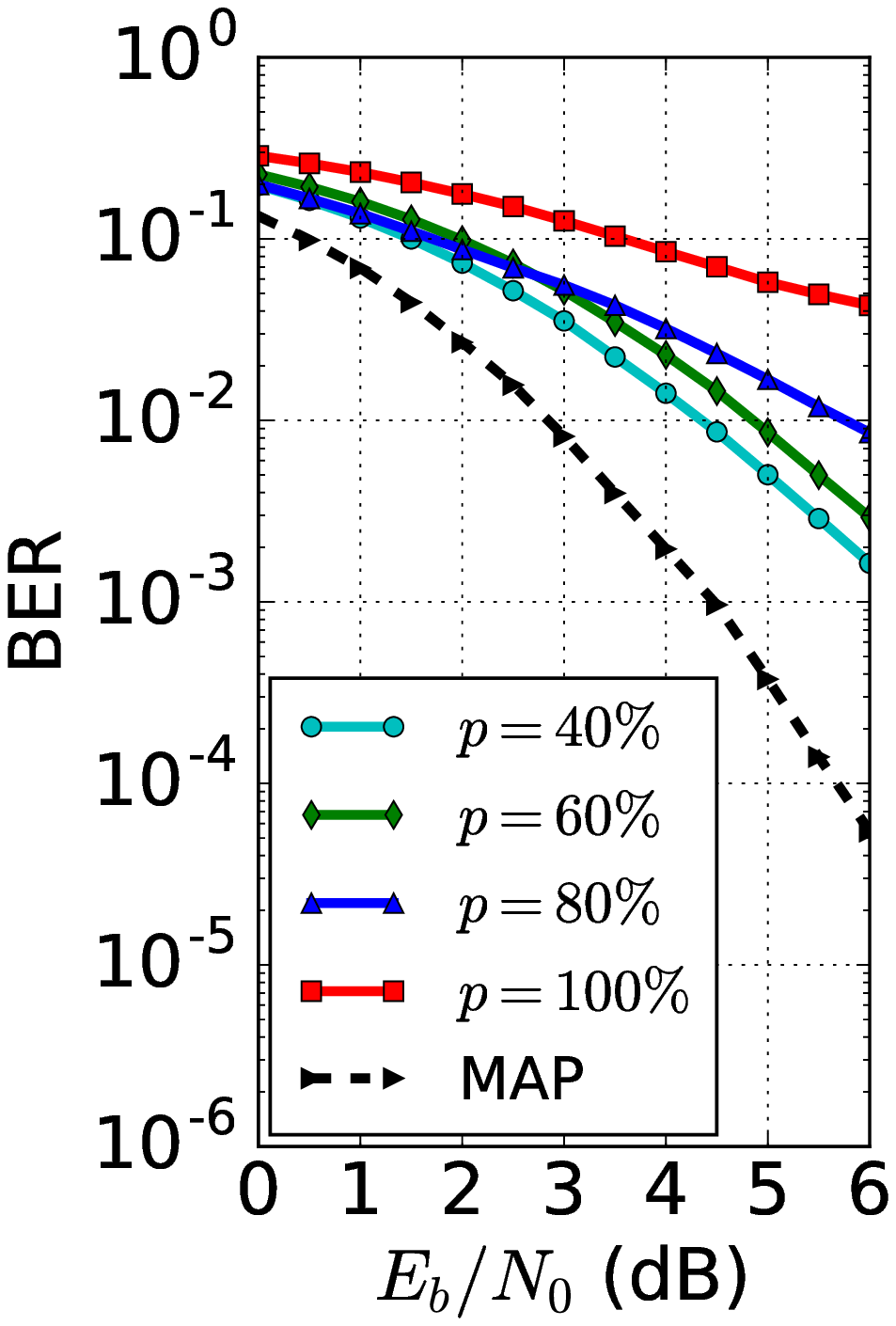}
	}
	\caption{The BER achieved by MLP, CNN and RNN with noise versus the testing SNR $E_b / N_0$ for $N = 32$ with training ratio $p = 40\%, 60\%, 80\%$ and $100\%$ and $M_{ep} = 10^5$.}
	\label{fig_32}
\end{figure}

Fig. \ref{fig_8} studies the case when $N = 8$, it can be observed that MLP, CNN and RNN can all achieve MAP performance when $p = 100 \%$, although they are overfitting when $p$ is under the $100\%$. Similarly, Fig. \ref{fig_16} studies the case when $N = 16$. For both MLP and CNN, we find that the four curves of different training ratio $p$ are very close and  have a certain gap compared with MAP curve, it indicated that both of MLP and CNN tend to be underfitting, although CNN is a little better than MLP. However, RNN still keeps overfitting who can achieve MAP performance when $p = 100\%$. From Fig. \ref{fig_32}, which studies the case when $N = 32$, we can conclude that both of MLP and CNN are in state of underfitting, while RNN appears the signs of underfitting as the case of MLP and CNN when $N = 16$.

Based on the results above, we summary that there exists a \emph{saturation length} for each type of deep neural network in NND, which is caused by their restricted learning abilities. When the codeword length $N$ is under the \emph{saturation length}, the complete encoding structure and noise characteristic can be well-learned only when the entire codebook is trained, i.e., $p = 100\%$, the NND will suffer from the problem of overfitting when $p$ is under $100\%$. When $N$ approaches the \emph{saturation length}, the performances with different training ratio $p$ are very close, and the NND tends to be underfitting with the training ratio $p$ increasing. When $N$ exceeds the \emph{saturation length}, the NND is thoroughly underfitting no matter what the $p$ is. For the proposed neural network architectures in this paper, we can conclude that the \emph{saturation lengths} of MLP and CNN are both $16$ despite that the learning ability of CNN is a little stronger than MLP, while the \emph{saturation length} of RNN is $32$.

\begin{figure}[!t] \centering
	\subfigure[] { \label{fig_backward}
		\includegraphics[scale = 0.22]{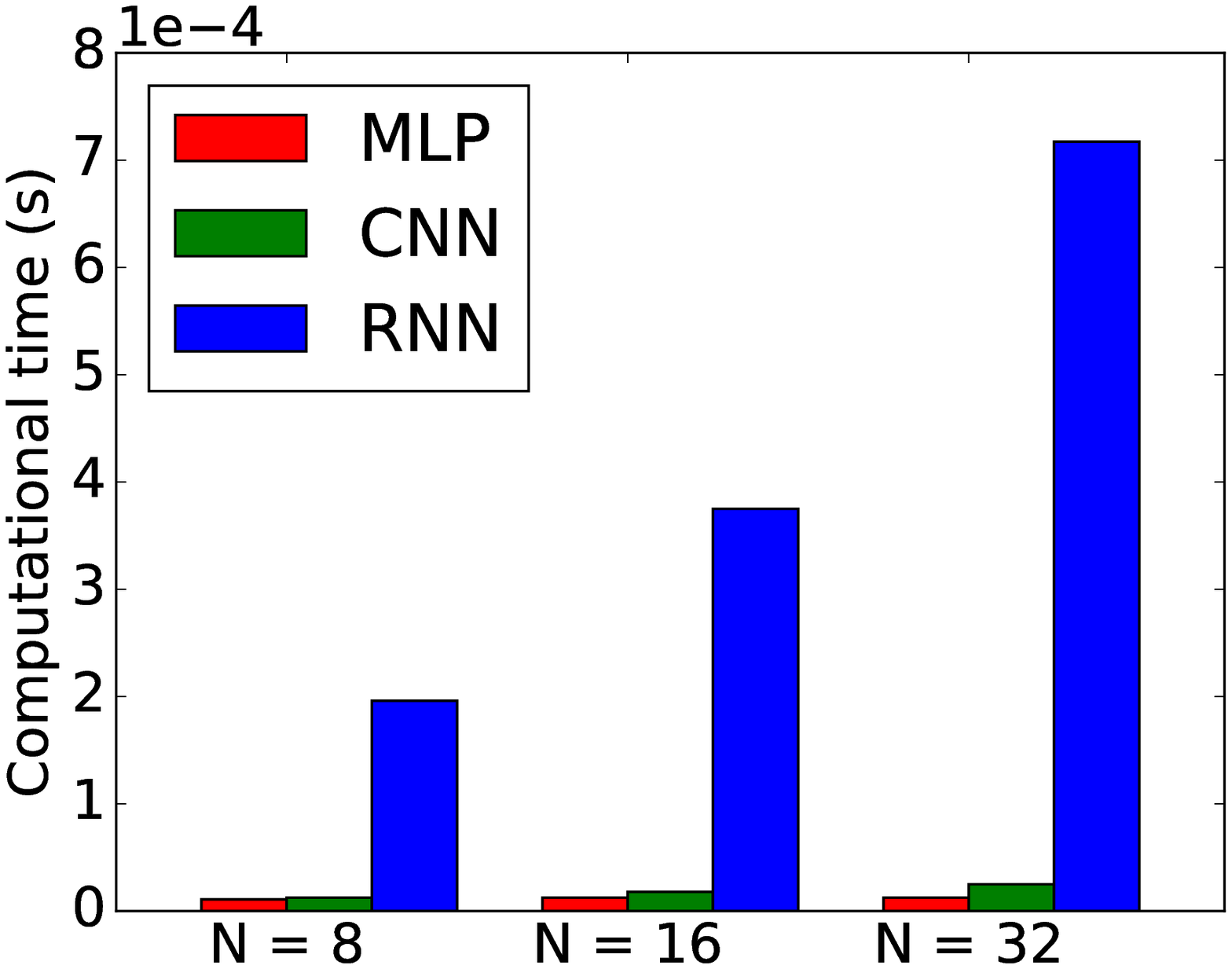}
	}
	\hspace{-0.8cm}
	\subfigure[] { \label{fig_forward}
		\includegraphics[scale = 0.22]{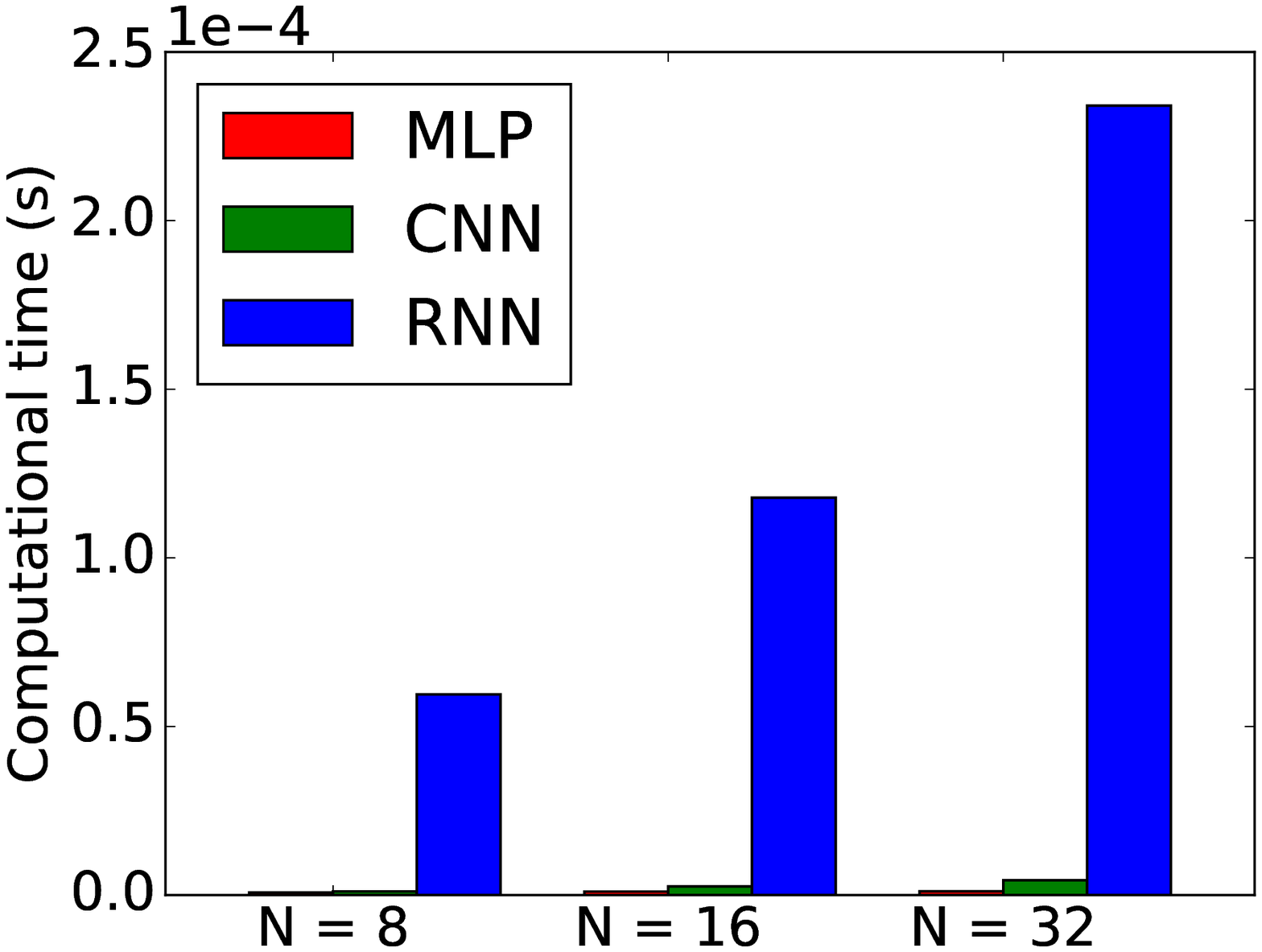}
	}
	\caption{The computational time of MLP, CNN and RNN versus the length of codeword $N$. (a) The backward propagation time for one training sample. (b) The forward propagation time for one testing sample.}
	\label{figure_time}
\end{figure}

We further investigate the computational time of MLP, CNN and RNN as shown in Fig. \ref{figure_time}. As mentioned before, we keep the same parameter magnitude for each neural network, however, the actual computational time is still very different due to their own specific structures. For both training and testing phase, the computational time of RNN is much higher than that of MLP and CNN, while the computational time of CNN is just a little bit higher than that of MLP. As such, we can conclude that the performances of CNN and MLP are very close, although CNN has better decoding performance but higher computational time in a small degree, and RNN can achieve the best decoding performance at the price of the highest computational time.

\section{Conclusion} \label{conclusion}

In this paper, we propose three types of NND, which build upon MLP, CNN and RNN. We compare the performance among these three deep neural networks through experiment, and find that RNN has the best decoding performance at the price of the highest computational time, and CNN has better decoding performance but higher computational time than MLP in a small degree. We find that the length of codeword influences the fitting of deep neural network, i.e., overfitting and underfitting. It is inferred that there exists a \emph{saturation length} for each type of neural network, which is caused by their restricted learning abilities. Reminding that the proposed structures of MLP, CNN and RNN are relatively simple and general, the focus of our future work is to design more complex structures to improve their \emph{saturation lengths} for the demand of longer codewords, for example, increasing neuron nodes for MLP, adding convolution layers for CNN or stacking more LSTM cells in each time step. In the meanwhile, the decoding performance with the \emph{saturation length} must approximate the MAP performance as far as possible.

\section*{Acknowledgement}
This work was supported in part by National Natural Science Foundation of China (Nos. 61725104, 61631003), Huawei Technologies Co.,Ltd (HF2017010003, YB2015040053, YB2013120029), and Education Foundation of Zhejiang University. 

\bibliographystyle{IEEEtran}

\end{document}